\documentclass[reprint,amsmath,amssymb,aps, prl]{revtex4-1}
\pdfoutput=1
\usepackage{graphicx}
\usepackage{bm}
\usepackage{verbatim}
\usepackage{multirow}
\usepackage{array}
\usepackage{threeparttable}
\usepackage{hyperref}
\usepackage{fancyhdr}

\begin{document}
\preprint{APS/123-QED}
\title{OnionNet-2: A Convolutional Neural Network Model for Predicting Protein-Ligand Binding Affinity based on Residue-Atom Contacting Shells}

\author{Zechen Wang}
\altaffiliation{School of Physics, Shandong University, Jinan, Shandong, China, 250100}
\author{Liangzhen Zheng}
\altaffiliation{Tencent AI Lab, Shenzhen, Guangdong, China, 518000}
\author{Yang Liu}
\altaffiliation{School of Physics, Shandong University, Jinan, Shandong, China, 250100}
\author{Yuanyuan Qu}
\altaffiliation{School of Physics, Shandong University, Jinan, Shandong, China, 250100}
\author{Yong-Qiang Li}
\altaffiliation{School of Physics, Shandong University, Jinan, Shandong, China, 250100}
\author{Mingwen Zhao}
\altaffiliation{School of Physics, Shandong University, Jinan, Shandong, China, 250100}
\author{Yuguang Mu}
\email{ygmu@ntu.edu.sg}
\altaffiliation{School of Biological Sciences, Nanyang Technological University, Singapore, 637551}
\author{Weifeng Li}
\email{lwf@sdu.edu.cn}
\altaffiliation{School of Physics, Shandong University, Jinan, Shandong, China, 250100}

\begin{abstract}
  One key task in virtual screening is to accurately predict the binding affinity ($\triangle$$G$) of protein-ligand complexes. Recently, deep learning (DL) has significantly increased the predicting accuracy of scoring functions due to the extraordinary ability of DL to extract useful features from raw data. Nevertheless, more efforts still need to be paid in many aspects, for the aim of increasing prediction accuracy and decreasing computational cost. In this study, we proposed a simple scoring function (called OnionNet-2) based on convolutional neural network to predict $\triangle$$G$. The protein-ligand interactions are characterized by the number of contacts between protein residues and ligand atoms in multiple distance shells. Compared to published models, the efficacy of OnionNet-2 is demonstrated to be the best for two widely used datasets CASF-2016 and CASF-2013 benchmarks. The OnionNet-2 model was further verified by non-experimental decoy structures from docking program and the CSAR NRC-HiQ data set (a high-quality data set provided by CSAR), which showed great success. Thus, our study provides a simple but efficient scoring function for predicting protein-ligand binding free energy.
\end{abstract}
\maketitle

\section{Introduction}
Protein-ligand binding is the basic of almost all processes in living organisms\cite{ref1} thus predicting binding affinity($\triangle$$G$) of protein-ligand complex becomes the research focus of bioinformatics and drug design. \cite{ref2, ref3, ref4} Theoretically, molecular dynamics (MD) simulations and free energy calculations (for instance, thermal integration method and free energy perturbation can provide accurate predictions of $\triangle$$G$ relying on extensive configurational sampling and calculation, leading to a large demand in computational cost.\cite{ref4, ref5, ref6, ref7} Therefore, developing simple, accurate and efficient scoring methods to predict protein-ligand binding will greatly accelerate the drug design process.\cite{ref8} To achieve this, several theoretical methods (scoring functions) have been proposed. Typically, the scoring functions are based on calculations of interactions between protein and ligand atoms.\cite{ref1, ref8, ref9, ref10} This includes quantum mechanics calculations, molecular dynamics simulations (electrostatic interaction, van der Waals interaction, hydrogen-bond and etc.), empirical-based interacting models.\cite{ref1, ref9, ref10, ref11}

In recent years, approaches based on machine learning (ML) have been applied in scoring functions and demonstrated great success.\cite{ref12, ref13, ref14, ref15} For instance, RF-Score\cite{ref16} and NNScore are two pioneering ML-based scoring functions.\cite{ref17} Compared with classical approaches, these ML-based methods allow higher flexibility in selecting configurational representations or features for protein and ligand. More importantly, these methods have been demonstrated to perform better and more effective than classical approaches.\cite{ref18, ref19} Recently, the deep learning (DL) approaches have provided alternative solution. Compared with ML, the DL models perform better at learning features from the raw data to extract the relationship between these features and labels.\cite{ref20, ref21, ref22} Thus, DL algorithms have been introduced to model the structure-activity relationships.\cite{ref23, ref24, ref25} One of the most popular methods of DL is the convolutional neural network (CNN), which is a multi-layer perceptron inspired by the neural network of living organisms.\cite{ref26}

Inspired by the great success of DL and CNN techniques, several models applying CNN to virtual screening and $\triangle$$G$ prediction have been reported.\cite{ref27, ref28, ref29, ref30, ref31, ref32, ref33, ref34, ref35} For example, \"{O}zt\"{u}rk and co-workers reported a DeepDTA model based on one-dimensional (1D) convolution, which took protein sequences and simplified molecular input line entry specification (SMILES) codes of ligand as inputs to predict drug-target $\triangle$$G$.\cite{ref30} Using 3D CNN model, two independent groups developed scoring functions, named Pafnucy\cite{ref31} and \emph{K}$_{deep}$\cite{ref29}, to model the complex in a cubic box centered on the geometric center of the ligand to predict the $\triangle$$G$ of protein-ligand complex. More interestingly, Russ et al. employed Graph-CNNs to automatically extract features from protein pocket and 2D ligand graphs, and demonstrated that the Graph-CNN framework can achieve superior performance without relying on protein-ligand complexes.\cite{ref34} Our group has proposed a 2D convolution-based predictor, called OnionNet, based on element-pair-specific contacts between ligands and protein atoms.\cite{ref32} As is shown, these DL and CNN based approaches, achieved higher accuracy in $\triangle$$G$ prediction than most traditional scoring functions, such as AutoDock,\cite{ref36, ref37} X-Score\cite{ref38} and KScore.\cite{ref39}

For DL scoring functions, how to treat with the high-dimensional structural information encoded in the 3D structures and convert to the low-dimensional features for ML (or DL) training is critical. For most structure-based ML/DL models, the features are usually derived from the atomic information of proteins and ligands, such as the element type and spatial coordinates of the atom and even other atomic properties.\cite{ref29, ref31}

We noticed that same elements in different residues have quite different physical and chemical properties, which might greatly affect the predicting performance of scoring functions. Considering that the twenty types of amino acids can be treated as intrinsic classifications of protein compounds which involve lay features of them, like polar, apolar, aromatic and etc. We anticipate that it may be more beneficial to encode protein as residues instead of atoms in developing DL scoring functions.

In this work, we proposed a simple OnionNet-2 -- a 2D CNN based regression model, which adopts the rotation-free residue-atom-specific contacts in multiple distance shells to describe the protein (residues) - ligand (atoms) interactions. We demonstrated that, our present method can significantly improve the prediction power by about 3.7\% than previous models, thus providing an efficient and accurate approach for predicting protein-ligand interactions and uncover a new trend of using DL technique for massive biological structures training for drug design.

\section{Methods}

    \subsection{Descriptors}
    The features employed are the pair numbers of the specific residue (protein)-atom (ligand) combination in multiple distance shells. The minimum distances between any atom in the ligand and any residue of protein are treated as the representative distances. First, around each atom in the ligand, we defined N continuously packed shells. The thickness of each shell is $\delta$, except that the first shell is a sphere with a radius of $d_0$. The boundary K$_i$ of the $i$th shell is as follows
    \begin{center}
        0 $<$ K{$_i$} $<$ $d_0$, $\,$ $i$ = 1 \\
        $d_0$ + (i - 2)$\delta$ $\le$ K{$_i$} $<$ $d_0$ + (i - 1)$\delta$, $\,$ i $\ge$ 2
    \end{center}

    Meanwhile, we classified atoms in the ligand into eight types, namely C, H, O, N, P, S, HAL and DU, where HAL represents the halogen elements (F, Cl, Br and I), and DU represents the element types excluded in these seven types.
    \begin{center}
    T$_e$ $\in$ \{C, H, O, N, P, S, HAL, DU\}
    \end{center}

    When pre-processing the structure file, water and ions were treated explicitly because crystal water molecules and ions could affect the protein-ligand binding.\cite{ref40, ref41} In addition to the twenty standard residues, we added an expanded type named ``OTH'' to represent water, ions and any other non-standard residues.
    \begin{center}
    T$_r$ $\in$ \{GLY, ALA, VAL, LEU, ILE, PRO, PHE, TYR, TRP, SER,
    THR, CYS, MET, ASN, GLN, ASP, GLU, LYS, ARG, HIS, OTH \}
    \end{center}

    \begin{figure}[htp]
      \centering
      \includegraphics[width=9cm, height=4.8cm]{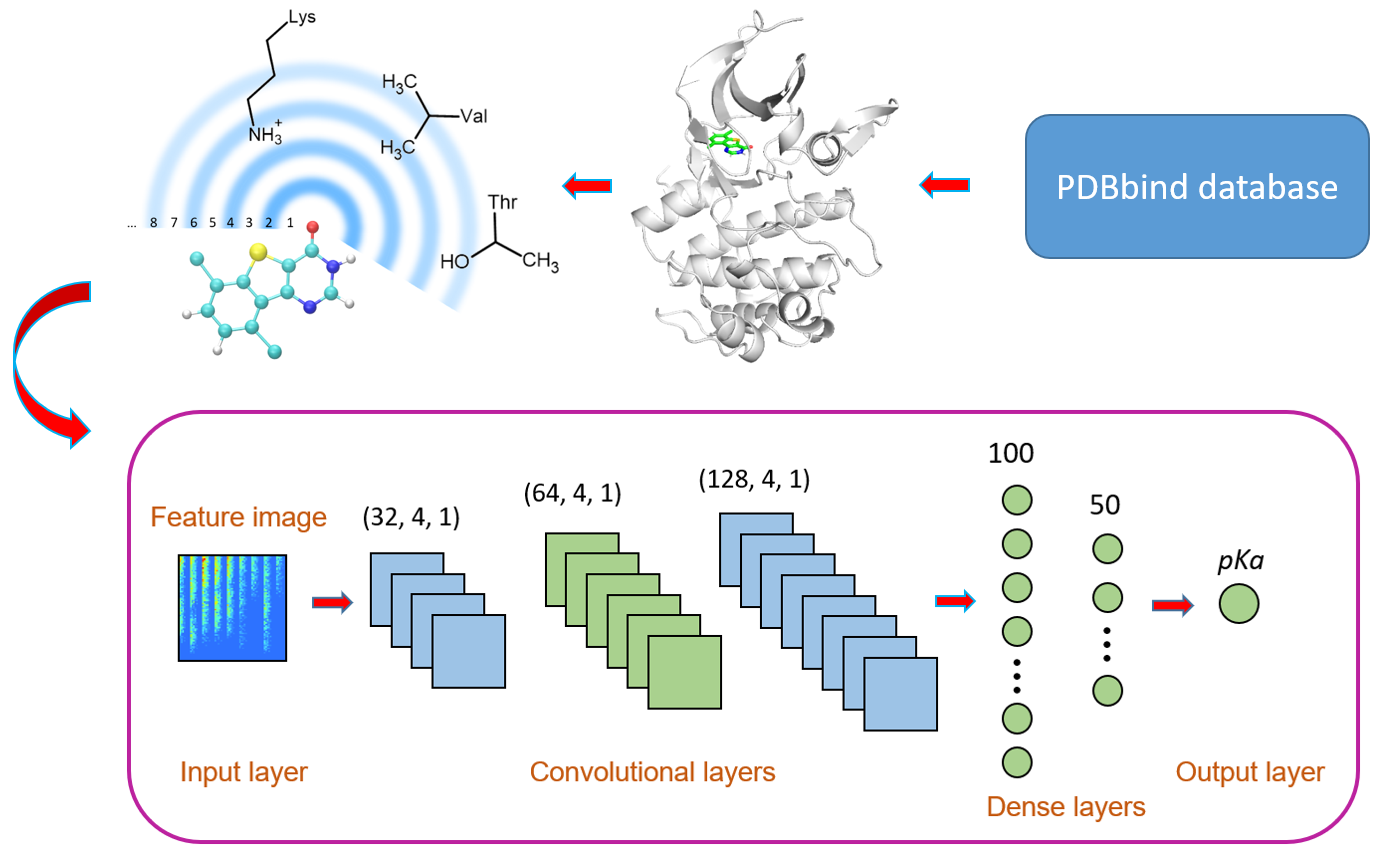}
      \caption{Workflow of OnionNet-2.}
    \end{figure}

    It is worth mention that the residue-atom distance is defined as the distance between the atom in the ligand and the nearest heavy atom in the residue. A 2D visual representation is depicted at the upper left of Fig. 1. For any shell, the number of contacts for each residue-atom pair is calculated and used as a feature. Each shell has 8$\times$21=168 residue-atom combinations, which means that there are 168 features for a shell. Thus, if the total number of shells is N, 168$\times$N features will be generated.
    \begin{equation}
    n_{T_r,T_e}=\sum_{r=1}^R\sum_{e=1}^Ec_{r,e}^i
    \end{equation}
    \begin{equation}
    c_{r,e}^i = \left\{
    \begin{aligned}
    \begin{gathered}
    1, (i-2)\delta + d_0 \leq d_{r,e} < (i-1)\delta + d_0 \\
    0, otherwise
    \end{gathered}
    \end{aligned}
    \right.
    \end{equation}

    Here, $R$ is the total number of residues in the protein, and $E$ is the total number of atoms in the ligand. The $d_{r,e}$ is the minimum distance between the residue $r$ in the protein and the atom $e$ in the ligand, and $n_{T_r,T_e}^i$ is the number of contacts of the specific residue-element combination in the ith shell. The $c_{r,e}^i$ is 1 when $(i - 2)\delta + d_0 \leq d_{r,e} < (i - 1)\delta + d_0$, otherwise $c_{r,e}^i$ is 0. Following our previous study,\cite{ref32} we used $d_0 = 1$ {\AA} and $\delta = 0.5$ {\AA}. Interestingly such shell-like, or radial, representations of protein environments, have been demonstrated to be superior features in protein function prediction.\cite{ref35} The preparation of datasets and the CNN architecture\cite{ref42, ref43} can be found in \href{run:./supporting/supplement.pdf}{SI}.  The source code of OnionNet-2 is available at \url{https://github.com/zchwang/OnionNet-2/}.

    \subsection{Evaluation metrics}
    To evaluate the performance of the OnionNet-2,we adopted the loss function defined in the previous work.\cite{ref32}
    \begin{equation}
    loss = \alpha(1 - R) + (1 - \alpha) RMSE
    \end{equation}
    \begin{equation}
    R=\frac{\sum_{i}^n(x_i-\bar x)(y_i-\bar y)}{\sqrt{\sum_i^n(x_i-\bar x)^2\sqrt{\sum_i^n(y_i-\bar y)^2}}}
    \end{equation}

    where R and RMSE represent Pearson correlation coefficient(Eq. 3) and root-mean-squared error, respectively; x$_i$ is the predicted pKa for ith complex; y$_i$ is the experimental pKa of this complex; $\bar x$ and $\bar y$ are the averages of all predicted values and experimental values.\cite{ref44} The $\alpha(0 \leq \alpha \leq1)$ value is an adjustable factor for adjusting the weight with R and RMSE, which was finally set to 0.7. For each independent training task, we adopted early stopping (patience = 20, that is, if the change of the loss value in the validating set is less than 0.001 after 20 epochs, the training is terminated) and save the model that performed best on the validating set. For the prediction in each case, five independently trainings were conducted to obtain the predicted mean value.

\section{Results and Discussions}

    \subsection{The predictive power of OnionNet-2}

    \begin{figure}[htp]
      \centering
      \includegraphics[width=6.6cm, height=5.4cm]{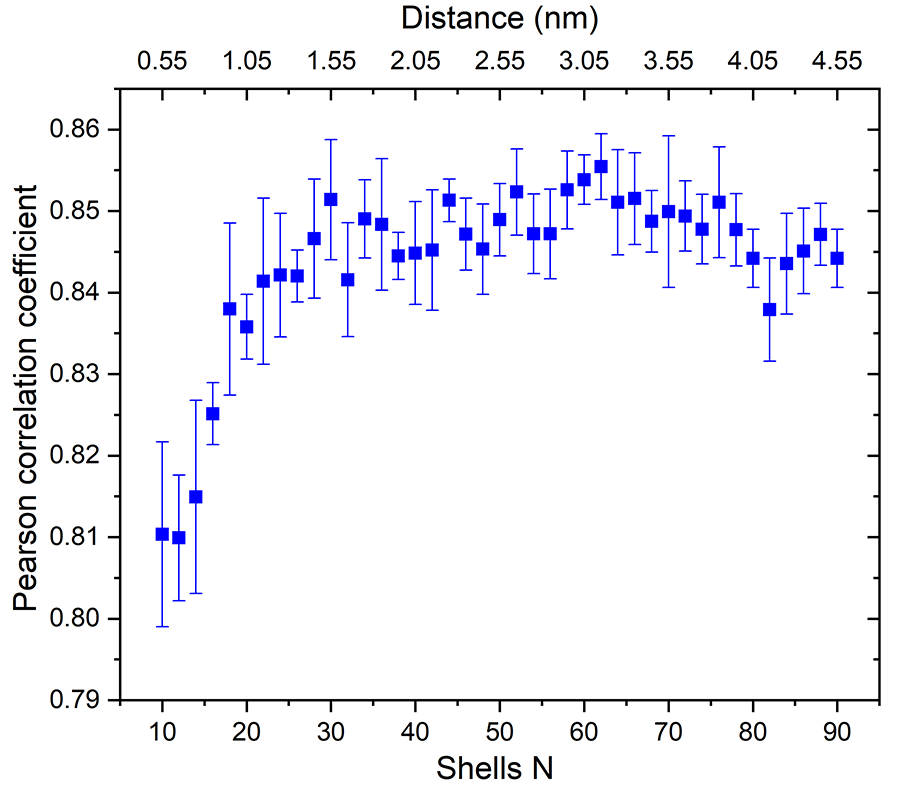}
      \caption{The Pearson correlation coefficient with respect to the shell number N for OnionNet-2 testing with core set v.2016. The bars indicate the standard deviations of the R values for five independent runs.}
    \end{figure}

    \begin{figure}[htp]
      \centering
      \includegraphics[width=8.5cm, height=8.5cm]{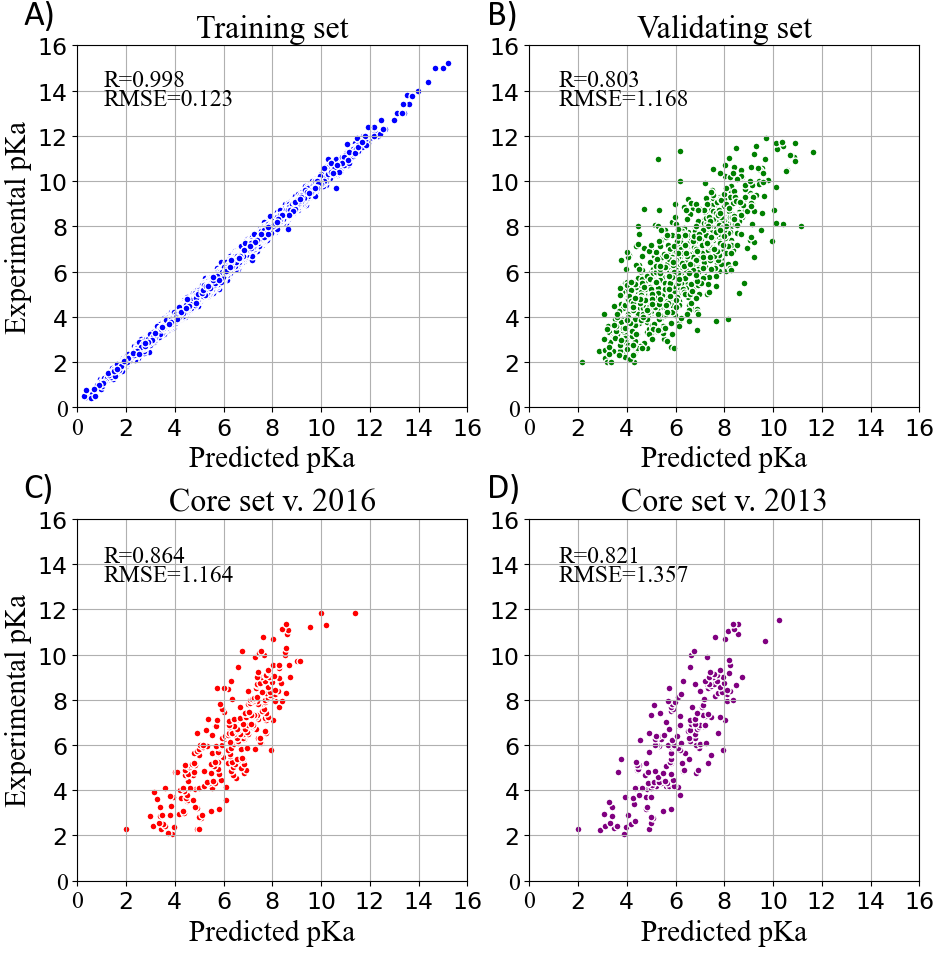}
      \caption{OnionNet-2 predicted pKa with respect to experimental determined pKa on (a) training set, (b) validating set, and two test sets of (c) core set v.2016 and (d) core set v.2013.}
    \end{figure}

    Firstly, we explored the effect of shell number N on the predictive capability of the OnionNet-2 model. A range of the total shell number $10 \leq N \leq 90$ was tested with interval of 2. According to our definitions of distance shell, this covers a separation between the residue and the atom from 0.55 nm to 4.55 nm. Fig. 2 depicts the trend of the R value to shell number N testing with core set v.2016\cite{ref44}. For N from 10 to 20, the R quickly increases as the total number of shells increases. This is expected because as the number of shells increases, the interactions between ligand and protein were gradually captured by the model. The R value reached the first peak for N is 30. This means that OnionNet-2 can achieve high prediction accuracy at a relatively low computational cost. Then, R fluctuates in a range of 0.01 until reaches the global maximum value when N = 62. Fig. 3 summarized the predicted mean pKa, with respect to experimental value, using N=62, on the training set (Fig. 3a), validating set (Fig. 3b) and two testing sets, core set v.2016 (Fig. 3c) and core set v.2013\cite{ref45, ref46, ref47} (Fig. 3d). It shows that the predicted pKa and experimental pKa are highly correlated for the two testing sets and validating set. After this point, R decreases when N increase. We attribute this to the enormous data that leads to the introduction of noise in the training. Unless otherwise specified, we adopted N of 62 in the following discussions. In addition, we also re-trained the model with two elder versions (v.2016 and v.2018) of the PDBbind database, and the R values of our re-trained models are almost the same (Fig. S1 and Table S1).

    The performance of some published scoring functions\cite{ref48, ref49} and OnionNet-2 tested on CASF-2016 and CASF-2013 are showed in Fig. 4 and Fig. S2, respectively. The corresponding R and RMSE (or SD) achieved by these representative scoring functions can be found in Table S2 in SI. Firstly, our OnionNet-2 model achieved highest R of 0.864 and RMSE of 1.164 with the core set v.2016, and R = 0.821 and RMSE = 1.357 with the core set v.2013. These were significantly higher than other scoring functions. The $2^{nd}$ best scoring function was AGL, which adopted the gradient boosting trees (GBTs) algorithm, focusing on multiscale weighted labeled algebraic subgraphs to characterize protein-ligand interactions.\cite{ref48} For two 3D convolution-based scoring functions \emph{K}$_{deep}$\cite{ref29} and Pafnucy\cite{ref31}, they adopted 3D voxel representation to model the protein-ligand complex and explicitly treated with physical properties of atoms such as hydrophobic, hydrogen-bond donor or acceptor and aromatic etc. into consideration. It is interesting to find that although we only employed the residue-atom contact to mimic the interactions between the protein and the ligand in OnionNet-2, the predicting power is higher. This further reveals that the selected features have a great impact on the predictive power of the CNN-based scoring functions. Secondly, as is expected, the introduction of ML/DL techniques into models has systematically enhanced the predicting accuracy.

    \begin{figure}[htp]
      \centering
      \includegraphics[width=8.8cm, height=4.4cm]{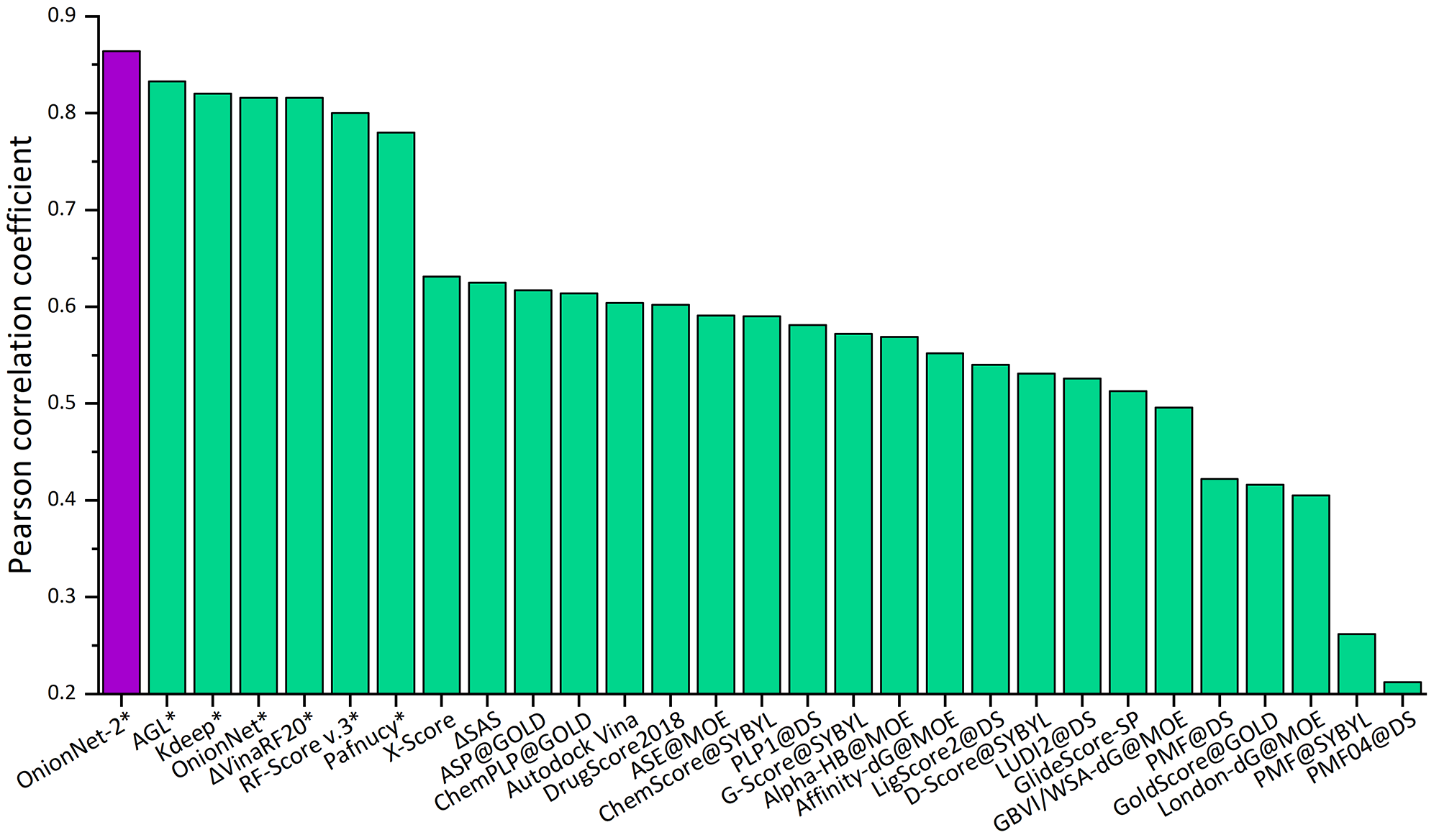}
      \caption{Pearson correlation coefficient of different scoring functions on CASF-2016 benchmark.}
    \end{figure}

    \subsection{Evaluation of the generalization ability of the model on different test sets}

    Generally, DL models display a good generalization behavior in practical applications.\cite{ref50} To verify the generalization ability of the OnionNet-2, the CSAR NRC-HiQ data set provided by CSAR\cite{ref51} was used as an additional test set in this study. This data set contains two subsets which contain 176 and 167 protein-ligand complexes respectively. For the two previous ML models, \emph{K}$_{deep}$ and RF-Score, the researchers used 55 and 49 complexes in two subsets respectively as test data.\cite{ref29} To provide a direct comparison with them, we adopted the same data for the OnionNet-2 test. It is worth mention that the two test subsets from the CSAR NRC-HiQ only have two common complexes with core set v.2013, namely 2jdy and 2qmj, and does not overlap with the training set, validation set and core set v.2016. The performance of \emph{K}$_{deep}$, RF-Score and OnionNet-2 on these two subsets are shown in Table \uppercase\expandafter{\romannumeral1}, and the scatter plots of the pKa predicted by OnionNet-2 with respect to experimental pKa can be found in Fig. S3 in SI.
    As expected, our model achieved a higher performance than \emph{K}$_{deep}$ and RF-score. For subset 1, the present OnionNet-2 achieved R of 0.89, which is considerably higher than that of \emph{K}$_{deep}$ (0.72) and RF-Score (0.78). This is also true for subset 2. Especially that, the R value of \emph{K}$_{deep}$ model is only 0.65 for subset 2, indicating weak predicting capability on these data. These results effectively demonstrated that OnionNet-2 has a good generalization ability.
    \begin{table}[htp]
      \begin{ruledtabular}
      \centering
      \caption{The performance of OnionNet-2, \emph{K}$_{deep}$ and RF-Score achieved on subsets from CSAR NRC-HiQ data set}
      \begin{tabular}{ccccc}
      {\,} & \multicolumn{2}{c}{Subset 1} & \multicolumn{2}{c}{Subset 2} \\
      {\,} & R & RMSE & R & RMSE \\
      \colrule
      OnionNet-2 & 0.89 & 1.50 & 0.87 & 1.21 \\
      \emph{K}$_{deep}$\cite{ref29} & 0.72 & 2.09 & 0.65 & 1.92 \\
      RF-Score\cite{ref29} & 0.78 & 1.99 & 0.75 & 1.66 \\
      \end{tabular}
      \end{ruledtabular}
    \end{table}
    \subsection{Evaluations on subsets of non-experimental decoy structures}
    As all the training and validating sets are composed of well-validated native structures in previous studies, it is largely unknown whether the DL method is capable to distinguish ``bad data'' that are incorporated in these integrated data sets, for instance, non-native binding poses. To verify this, we tested OnionNet-2 to deal with non-experimental structures (generated by docking programs). Technically, non-native binding poses (called decoys) were generated based on core set v.2016 complexes by AutoDock Vina.\cite{ref52, ref53} The detailed information of the generation of decoys can be found in the SI.\cite{ref54, ref55}

    The predicting accuracy was evaluated by calculating the RMSE between the predicted pKa of the decoy complex and the pKa of the corresponding native receptor-ligand complex which is shown in Fig. 5. It is clear that, the RMSE quickly increased with increasing RMSD. This is expected because decoys with larger RMSD result in more severe change of $\delta$G. These results reveal that OnionNet-2 can accurately respond to changes of the ligand binding poses and distinguish the native structure.
    \begin{figure}[htp]
      \centering
      \includegraphics[width=7cm, height=5.6cm]{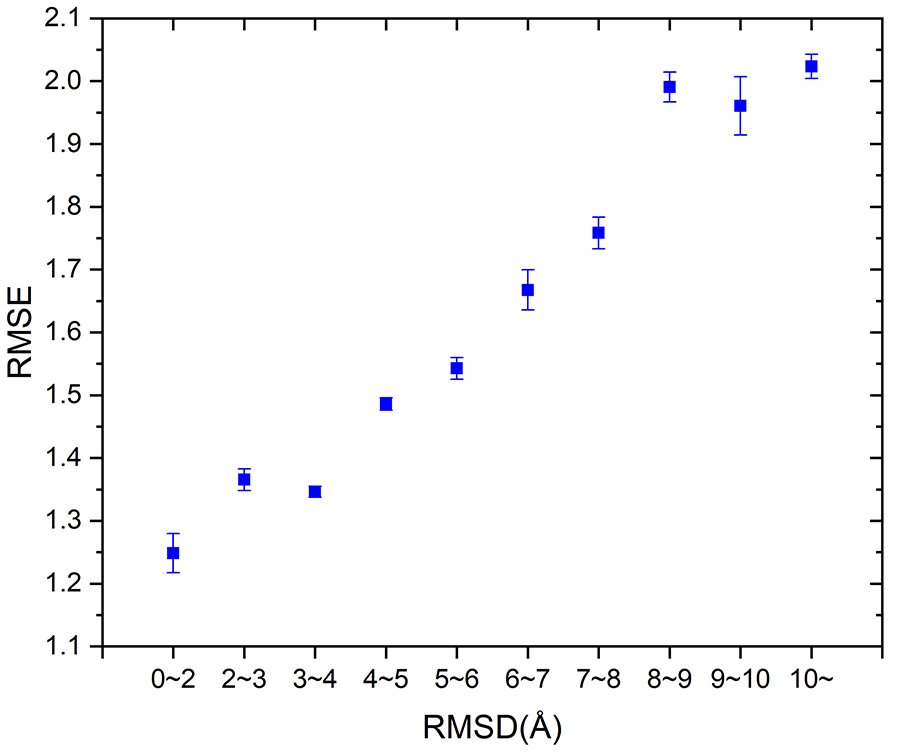}
      \caption{RMSE between the predicted pKa of the decoy complex and the pKa of the corresponding native receptor-ligand complex achieved by OnionNet-2 on different protein-decoy complexes subsets. The bars indicate the standard deviations of RMSE for five independent runs.}
    \end{figure}

    \begin{table*}[htp]
      \centering
      \caption{Pearson correlation coefficients achieved by OnionNet-2 and previous scoring functions on three series of subsets}
      \begin{threeparttable}
      \begin{ruledtabular}
      \begin{tabular}{cccccccccc}
      {\,} & \multicolumn{3}{c}{Subset H} & \multicolumn{3}{c}{Subset S} & \multicolumn{3}{c}{Subset V} \\
      {\,} & H1 & H2 & H3 & S1 & S2 & S3 & V1 & V2 & V3 \\
      \colrule
      OnionNet-2 & 0.872 & 0.866 & 0.856 & 0.868 & 0.839 & 0.869 & 0.856 & 0.774 & 0.866 \\
      $\triangle$$_{Vina}$RF$_{20}$\tnote{*} & 0.820 & 0.832 & 0.804 & 0.843 & 0.765 & 0.823 & 0.727 & 0.760 & 0.818 \\
      X-Score\tnote{*} & 0.698 & 0.570 & 0.661 & 0.743 & 0.536 & 0.572 & 0.437 & 0.622 & 0.579 \\
      X-Score$^{HS}$\tnote{*} & 0.711 & 0.565 & 0.647 & 0.748 & 0.557 & 0.546 & 0.433 & 0.630 & 0.559 \\
      $\triangle$SAS\tnote{*} & 0.641 & 0.643 & 0.589 & 0.746 & 0.572 & 0.494 & 0.480 & 0.669 & 0.541 \\
      X-Score$^{HP}$\tnote{*} & 0.669 & 0.558 & 0.672 & 0.740 & 0.510 & 0.575 & 0.450 & 0.575 & 0.580 \\
      AutoDock Vina\tnote{*} & 0.626 & 0.586 & 0.641 & 0.745 & 0.521 & 0.507 & 0.484 & 0.564 & 0.486 \\
      \end{tabular}
      \end{ruledtabular}
      \begin{tablenotes}
      \footnotesize
      \item[*] Results of the last six rows taken from Su et al.\cite{ref44}
      \end{tablenotes}
      \end{threeparttable}
    \end{table*}

    \subsection{Effects of hydrophobic scale, buried solvent-accessible area and excluded volume inside the binding pockets on the prediction accuracy}

    Principally, the physical interactions between protein and ligand determine the $\triangle$$G$. The dominating factors for overall $\triangle$$G$ involve electrostatic interactions, van der Waals interactions, hydrogen bonds, hydration/de-hydration during complexation. However, such mechanistic interactions were not directly input into DL features. At molecular level, these involves the size and shape of the binding pocket, and the nature of residues around the binding pocket which determine its physicochemical characteristics.\cite{ref56} Whether DL models can accurately represent the structural specificity of the binding pocket is poorly documented.

    The entire CASF-2016 test set can be divided into three subsets by each of three descriptors according to physical classifications of the binding pocket on the target protein.\cite{ref44} The three descriptors include H-scale (hydrophobic scale of the binding pocket), $\triangle SAS$ (buried percentage of the solvent-accessible area of the ligand after binding) and $\triangle VOL$ (excluded volume inside the binding pocket after ligand binding). Protein-ligand complexes in CASF-2016 were grouped into 57 clusters, and the authors sorted all 57 clusters in ascending order by each descriptor. Then, these complex clusters were divided into three subsets according to the chosen descriptor, labeled as H1, H2 and H3 or S1, S2 and S3 or V1, V2 and V3. These subsets were used as validations of our OnionNet-2 model. As comparison, previous scoring functions were also tested on these three sets of subsets by Su et al.\cite{ref44}, and the results are summarized in Table \uppercase\expandafter{\romannumeral2}.

    As can be seen in Table \uppercase\expandafter{\romannumeral2}, OnionNet-2 achieved higher prediction accuracy compared with other soring functions when tested on H-, S- and V-series subsets. This indicates that the feature based on the contact number of residue-atom pairs in multiple shell is capable of capturing the hydrophobic scale of the binding pocket. The number of contacts in different shells (specifically the shells within the binding pocket) may be able to reflect the buried solvent-accessible surface area and the excluded volume of the ligand.

    We noticed that, compared to other subsets, the R value of OnionNet-2 on V2 subset is clearly lower than other subsets (nevertheless it is still high than other scoring functions). This may indicate that our model is less sensitive to medium-sized binding pockets. Thus it may be still challenging for current scoring functions to recognize the size and shape of the binding pocket.

    Furthermore, we plotted the detailed scatter plots of predicted pKa and experimental pKa in Fig. S4 according to the specific H, S and V range. It is interesting to find almost no dependence of pKa with the values of H, S or V. Thus we speculate that a more realistic descriptor for the ligand characteristic in the binding pocket is essential to guide the protein-ligand $\triangle$$G$ prediction.

\section{Conclusion}

To summarize, a 2D convolution-based CNN model, OnionNet-2, is proposed for prediction of the protein-ligand binding free energy. The contacting pair numbers between the protein residues and the ligand atoms were used as features for DL training. Using CASF-2013 and CASF-2016 as benchmarks, our model achieved the highest accuracy to predict $\triangle$$G$ than previous scoring functions. In addition, when employing different versions of PDBbind database for training, the performance of OnionNet-2 is nearly the same. The generalization ability of the model was verified by the CSAR NRC-HiQ data set and decoy structures. Our result also indicates that OnionNet-2 has the capability to recognize these physical natures (in detail, hydrophobic scale of the binding pocket, buried percentage of the solvent-accessible area of the ligand upon binding and excluded volume inside the binding pocket upon ligand binding) of the ligand-binding pocket interactions.

\begin{acknowledgments}
This work is supported by the Natural Science Foundation of Shandong Province (ZR2020JQ04), National Natural Science Foundation of China (11874238) and Singapore MOE Tier 1 Grant RG146/17.
\end{acknowledgments}

\end{document}


Supporting Information\\
\title{OnionNet-2: A Convolutional Neural Network Model for Predicting Protein-Ligand Binding Affinity based on Residue-Atom Contacting Shells}

\author{Zechen Wang}
\altaffiliation{School of Physics, Shandong University, Jinan, Shandong, China, 250100}
\author{Liangzhen Zheng}
\altaffiliation{Tencent AI Lab, Shenzhen, Guangdong, China, 518000}
\author{Yang Liu}
\altaffiliation{School of Physics, Shandong University, Jinan, Shandong, China, 250100}
\author{Yuanyuan Qu}
\altaffiliation{School of Physics, Shandong University, Jinan, Shandong, China, 250100}
\author{Yong-Qiang Li}
\altaffiliation{School of Physics, Shandong University, Jinan, Shandong, China, 250100}
\author{Mingwen Zhao}
\altaffiliation{School of Physics, Shandong University, Jinan, Shandong, China, 250100}
\author{Yuguang Mu}
\email{ygmu@ntu.edu.sg}
\altaffiliation{School of Biological Sciences, Nanyang Technological University, Singapore, 637551}
\author{Weifeng Li}
\email{lwf@sdu.edu.cn}
\altaffiliation{School of Physics, Shandong University, Jinan, Shandong, China, 250100}

\maketitle

\setcounter{equation}{0}
\setcounter{figure}{0}
\setcounter{table}{0}
\setcounter{page}{1}
\makeatletter
\renewcommand{\theequation}{S\arabic{equation}}
\renewcommand{\thefigure}{S\arabic{figure}}
\renewcommand{\thetable}{S\arabic{table}}
\section{Part \uppercase\expandafter{\romannumeral1}. Preparation of Dataset and Architecture}
\subsection{Preparation of training set, validating set and testing sets}
We mainly used the protein-ligand complexes of PDBbind database v.2019 (\url{http://www.pdbbind-cn.org/}) for training. This database consists of two overlapping subsets, the general set and the refined set. The general set includes all available complexes and the refined set comprises protein-ligand complexes with high-quality structure and binding information selected from the general set. For each structure of the protein-ligand complex, the corresponding binding affinity is represented by the negative logarithms (pKa) of the dissociation constants ($Kd$), inhibition constants ($Ki$) or half inhibition concentrations ($IC50$). In order to evaluate the predictive ability and compare with other scoring functions, OnionNet-2 was evaluated on the CASF-2016 test set (core set v.2016)\cite{ref1} and CASF-2013 test set (core set v.2013)\cite{ref2, ref3, ref4}. It should be noted that the CASF-2016 test set is the latest update of CASF-2016, which contains 285 high-quality complexes. While for core set v.2013, it is a subset of the PDBbind database v.2013, consisting of 195 protein-ligand complexes classified in 65 clusters with binding constants spanning nearly 10 orders of magnitude. Besides, a data set called CSAR NRC-HiQ, consisting of two subsets containing 176 and 167 complexes respectively,\cite{ref5} was employed as a third test set. For the previous models of Kdeep and RF-score, 55 and 49 complexes in two subsets were used as test data.\cite{ref6} To provide a direct comparison with \emph{K}$_{deep}$ and RF-score, we adopted the same data for the OnionNet-2 test. In order to perform normal training and testing, it is necessary to redistribute remaining complexes in PDBbind database v.2019. First, we excluded the complexes contained in three test sets from PDBbind database v.2019 (general set and refined set). Then, as a common practice (Reference: Pafnucy\cite{ref7} and OnionNet\cite{ref8}), 1000 complexes were randomly sampled from v.2019 refined set (after filtering all complexes used in the test sets described previously) as the validating set. Finally, the remaining complexes (that is, the complexes that are not included in the three test sets and validating set) were adopted for the training set. This ensures that there is no overlapping protein-ligand complex in the training set, validating set and test sets.

\subsection{Architecture}
We adopted a CNN model based on 2D convolution to learn the relationship between the contact features and the $\triangle$$G$. The model was constructed using the Keras package in tensorflow.\cite{ref9} The workflow architecture is shown in Fig. 1.

The raw data is pre-processed before input into the CNN model. Here, the features are standardized through the scikit-learn package.\cite{ref10} Three convolutional layers, with 32, 64 and 128 filters respectively were used and the filter sizes were all set as 4, with strides as 1. After preliminary tests, two fully connected layers with 100 and 50 neurons are used before the output layer, which is capable of capturing the nonlinear relationship between the features and the pKa values.

To further increase the nonlinear ability of the model, a rectified linear unit (RELU) layer was added after each convolutional layer and fully connected layer. Also, a batch normalization layer was used after the fully connected layer. The stochastic gradient descent (SGD) optimizer was adopted and the learning rate was set as 0.001. To reduce overfitting, L2 regularization with weight decay $\lambda$ = 0.01 was used after each fully connected layer. The number of samples processed per batch is 64.

\section{Part \uppercase\expandafter{\romannumeral2}. The statistical information of the training set and validating set when using different versions of PDBbind database}

Generally, the size of the datasets has an impact on the predictive ability of the DL model.\cite{ref11, ref12} To make a fair comparison with previous scoring functions, we re-trained the model with two elder versions (v.2016 and v.2018) of the PDBbind database, and the number of samples in the training set and validating set is shown in Table S1. In addition, we only re-trained the models using N = 58, 60 and 62. The performance on core set v.2013 and core set v.2016 are summarized in Fig. 4. It is clear that, although the three versions of PDBbind database differed greatly in size, the R values of our re-trained models are almost the same. This suggests that the difference between these three databases (in detail, PDBbind database v.2016, v.2018 and v.2019) has a rather limited impact on our OnionNet-2 model.

    \begin{figure}[htp]
      \centering
      \includegraphics[width=9cm, height=4cm]{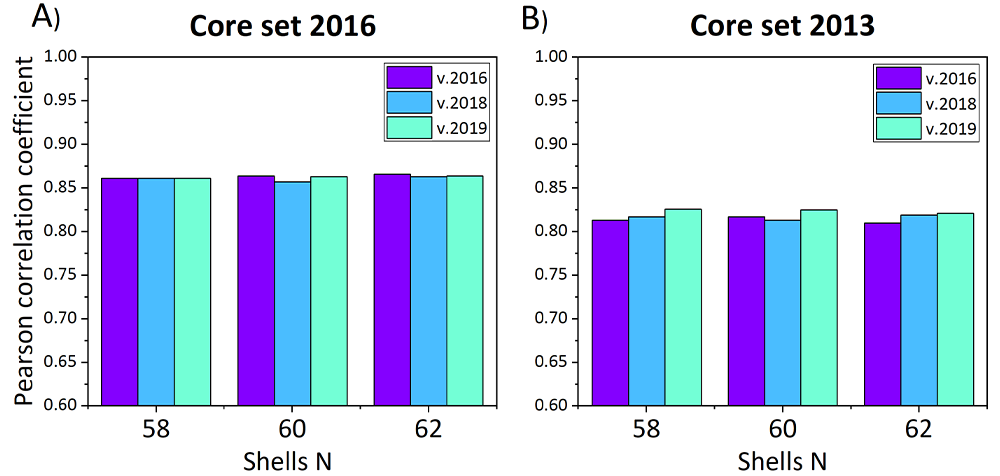}
      \caption{Pearson correlation coefficients achieved by OnionNet-2 on (A) core set v.2016 and (B) core set v.2013 when different versions of PDBbind database are used for training (the total number of shells is set to 58, 60 and 62).}
    \end{figure}

    \begin{table}[htp]
      \begin{ruledtabular}
      \centering
      \caption{The number of samples in training set and validating set for three versions of PDBbind database.}
      \begin{tabular}{ccc}
      {\,} & Training set & Validating set \\
      \colrule
      v.2016 & 11820 & 1000 \\
      v.2018 & 14686 & 1000 \\
      v.2019 & 16626 & 1000 \\
      \end{tabular}
      \end{ruledtabular}
    \end{table}

\section{Part \uppercase\expandafter{\romannumeral3}. Comparison of OnionNet-2 with some representative scoring functions}
    \begin{table}[htp]
      \begin{ruledtabular}
      \centering
      \caption{Comparison of the predictive power of scoring functions on the core set v.2016 and v.2013}
      \begin{tabular}{ccccc}
      {\,} & \multicolumn{2}{c}{CASF-2016} & \multicolumn{2}{c}{CASF-2013} \\
      \colrule
      Scoring functions & R & RMSE & R & SD \\
      OnionNet-2 & 0.864 & 1.164 & 0.821 & 1.29 \\
      AGL\cite{ref13} & 0.833 & 1.271 & 0.792 & 1.45 \\
      \emph{K}$_{deep}$\cite{ref6} & 0.82 & 1.27 & {} & {} \\
      OnionNet\cite{ref8} & 0.816 & 1.278 & 0.78 & 1.45 \\
      RF-Score-v3\cite{ref7} & 0.80 & 1.39 & 0.74 & 1.51 \\
      Pafnucy\cite{ref7} & 0.78 & 1.42 & 0.70 & 1.61 \\
      kNN-Score\cite{ref14} & {} & {} & 0.672 & 1.65 \\
      X-Score\cite{ref14} & {} & {} & 0.614 & 1.78 \\
      ChemScore\cite{ref14} & {} & {} & 0.592 & 1.82 \\
      AutoDock Vina\cite{ref15} & {} & {} & 0.54 & 1.90 \\
      AutoDock\cite{ref15} & {} & {} & 0.54 & 1.91 \\
      \end{tabular}
      \end{ruledtabular}
    \end{table}

    \begin{figure}[htp]
      \centering
      \includegraphics[width=9cm, height=4.8cm]{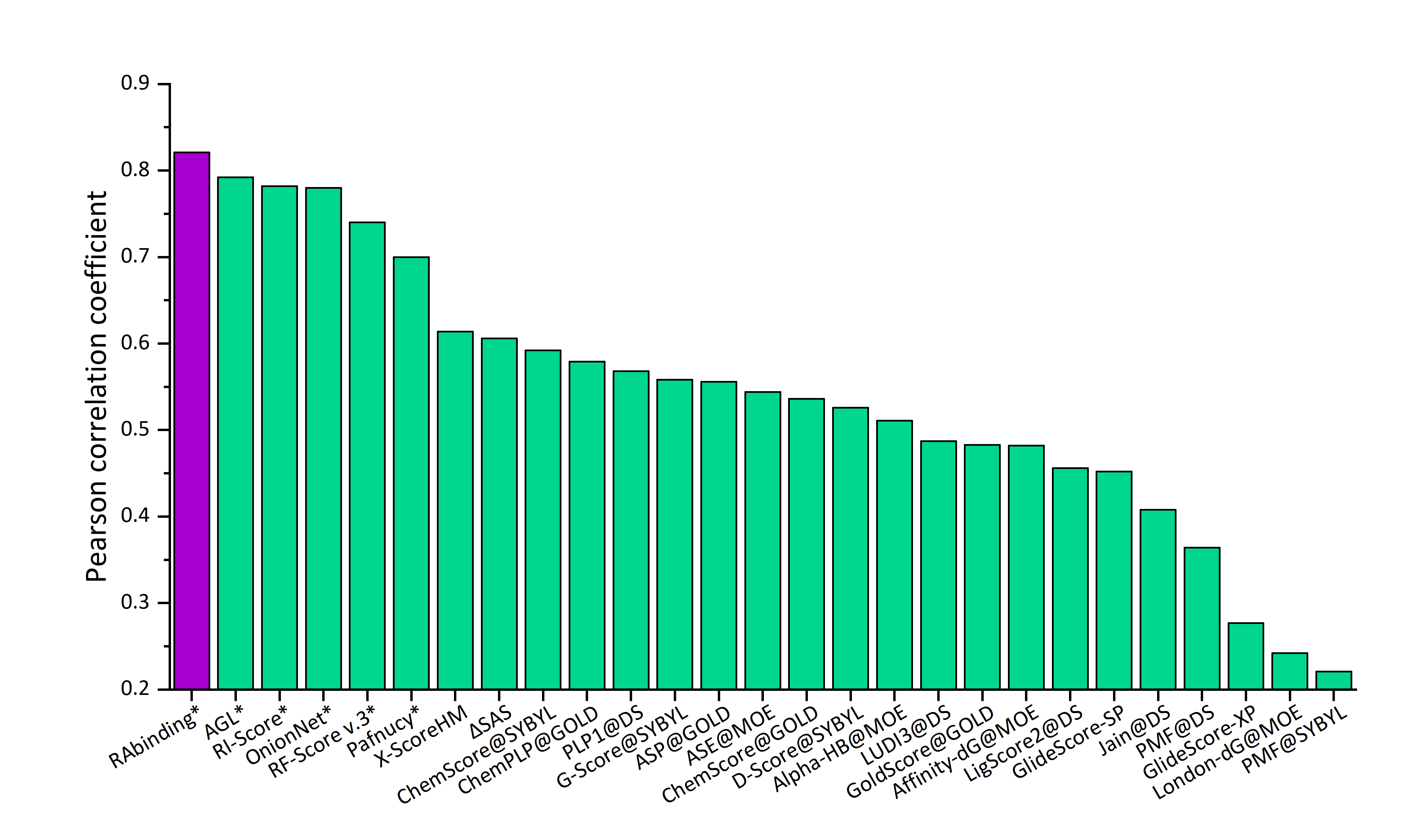}
      \caption{Pearson correlation coefficient of different scoring functions on two benchmarks CASF-2013.}
    \end{figure}

    \begin{figure}[h!]
      \centering
      \includegraphics[width=9cm, height=4.4cm]{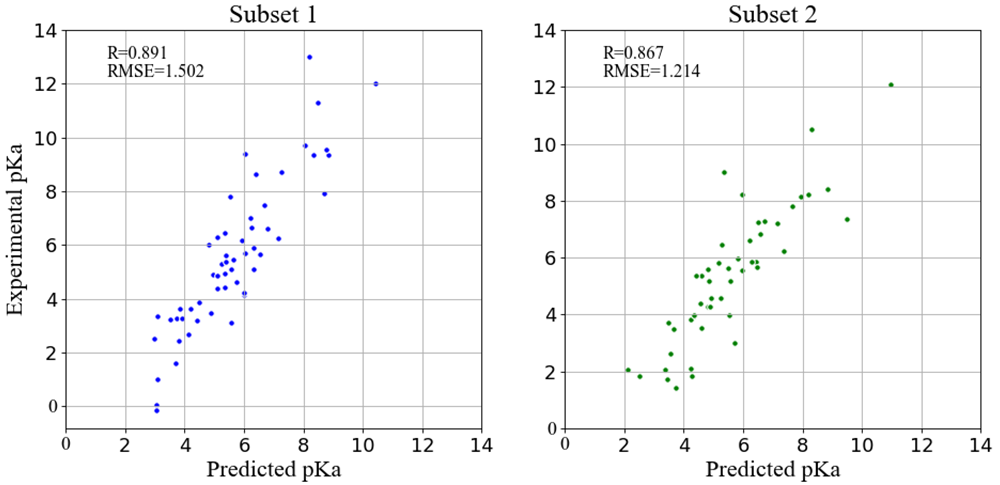}
      \caption{The result achieved by OnionNet-2 on CSAR-HiQ subsets.}
    \end{figure}

\section{Part \uppercase\expandafter{\romannumeral4}.The detailed information of the generation of decoys}

In CASF-2016 benchmark, the similarity between two binding poses is measured by the root-mean-square deviation (RMSD) value.\cite{ref4} Following previous studies by Allen et al.,\cite{ref16} we adopted the Hungarian algorithm to calculate RMSD between decoy ligand and native structure which is implemented in spyrmsd.\cite{ref17} In this study, we used AutoDock Vina to generate non-native binding pose (called decoys). The sampling space is a 27 {\AA} $\times$ 27 {\AA} $\times$ 27 {\AA} cubic, which is centered on the geometric center of the native ligand binding pose. The exhaustiveness value was set to 12. It is worth noting that AutoDock Vina ignores the effects of water molecules and ions when calculating protein-ligand binding energy. In addition, changes in the ligand binding poses will be accompanied by changes in the surrounding microenvironment. Therefore, the receptor-decoy complexes here do not include water molecules and ions.

1.	For each receptor, up to 20 decoy ligands were generated by AutoDock Vina. The actual number may be less than 20 because of limited size and shape of the binding pocket in the target protein. For each decoy, the RMSD with respect to native structure was calculated.

2.	We used 10 RMSD intervals, [0 {\AA}, 2 {\AA}], [2 {\AA}, 3 {\AA}], [3{\AA}, 4{\AA}], ..., [9{\AA}, 10{\AA}] and [10{\AA}:].

3.	For all ligands in every interval, we selected the decoy with the smallest RMSD value to put into the corresponding subsets.

4.	Ten test subsets containing non-experimental complexes were used for OnionNet-2 training.

\section{Part \uppercase\expandafter{\romannumeral5}. OnionNet-2 predicted pKa with respect to experimental determined pKa on three series subsets}
\begin{figure}[htp!]
  \centering
  \includegraphics[width=10cm, height=3cm]{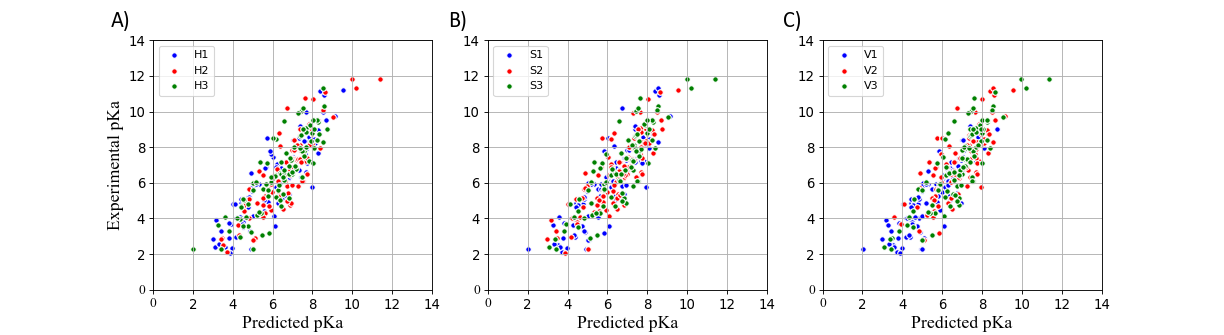}
  \caption{OnionNet-2 predicted pKa with respect to experimental determined pKa on (A) H-, (B) S- and (C) V series subsets.}
\end{figure}